\documentclass[showpacs,preprintnumbers,amsmath,amssymb,12pt]{revtex4}
\usepackage{graphicx}
\usepackage{dcolumn}
\usepackage{xcolor}
\usepackage{bm}
\usepackage{amsfonts}
\usepackage{mathrsfs}
\begin{document}

\title{Universality of Quasinormal-Mode Shifts from Small Nonlocal Effective Couplings}
\author{Anisur Rahaman}
\email{manisurn@gmail.com; anisur@associates.iucaa.in}
\affiliation{Durgapur Government College, Durgapur-713214, India}

\date{\today}
\begin{abstract}
We investigate perturbative quasinormal-mode (QNM) shifts of black holes arising from fractional, nonlocal modifications to the wave operator. Starting from a scalar master equation corrected by a small fractional Laplacian term $(-\Delta)^{s}$ with $0<s<1$, we derive an analytic expression for the complex frequency shift at first order in the nonlocal coupling $\varepsilon$. Evaluation of the fractional operator in both coordinate and momentum representations reveals a universal scaling law $\delta\omega/\omega \propto \varepsilon/M^{2s}$, largely independent of the field spin, with an additional $\ell^{2s}$ enhancement in the eikonal regime $\ell \gg 1$. Applying the formalism to Schwarzschild, slowly rotating Kerr, Hayward regular, and LQG-corrected black holes, we demonstrate that the leading-order fractional QNM shift is universal, with geometric details entering only through overlap integrals of the mode functions. This universality provides a model-independent signature of nonlocality in strong-gravity ringdown spectra and offers a potential observational window into quantum-gravity-inspired modifications.
\end{abstract}

\maketitle

\section{Introduction}
Quasinormal modes (QNMs) of black holes encode the characteristic damped oscillations of perturbations in the black-hole spacetime. These modes depend solely on the background geometry and the field equations governing the perturbations~\cite{Berti:2009kk,Konoplya:2011qq,Kokkotas:1999bd,Nollert:1999ji}. The detection of gravitational waves from binary black-hole mergers by LIGO and Virgo has opened a new avenue for using the ringdown spectrum as a precision probe of strong-gravity physics~\cite{ChenPark2021,BryantYagi2021,Abbott:2016blz,Isi:2019aib,Giesler:2019uxc}.

While general relativity (GR) provides a remarkably accurate description of astrophysical black holes, it is widely expected to be modified at high energies or short distances. Effective-field-theory approaches introduce higher-derivative curvature corrections~\cite{Cano:2020cao,Cardoso:2018ptl}. Quantum-gravity candidates, such as string theory and loop quantum gravity, often predict nonlocality, minimal length scales, or modified dispersion relations~\cite{Modesto:2011kw,Buoninfante:2018mre,Antoniou2025Quadratic,Calcagni:2010bj,Briscese:2018bny,Conroy:2015wfa}. These modifications typically appear as corrections to the effective wave operator governing linear perturbations, resulting in small but potentially measurable shifts in the QNM frequencies~\cite{Tattersall:2018axd,Franchini:2021bpt}.

A particularly intriguing class of corrections involves fractional, nonlocal operators, where the conventional Laplacian $\Delta$ is replaced by its fractional counterpart $(-\Delta)^s$ with $0 < s < 1$. Fractional Laplacians naturally arise in nonlocal field theories, effective descriptions of ultraviolet-complete gravity, and models incorporating minimal length scales~\cite{Buoninfante:2018mre,Roy2022Fractional,Barrow2017MultiFractional,Calcagni:2017sdq,Cembranos:2020tix}. These operators induce a scale-dependent modification of the perturbation dispersion relation, affecting both the oscillation frequency and damping rate of QNMs~\cite{Spallucci:2017aod,Modesto:2017sdr}.

In this work, we develop an analytical and model-independent framework to evaluate first-order shifts in black-hole QNMs due to fractional, nonlocal perturbations. We derive explicit expressions for the frequency shifts in both coordinate and momentum space and apply the formalism to several representative black-hole spacetimes, including Schwarzschild, slowly rotating Kerr, Hayward regular, and LQG-corrected black holes. We demonstrate a remarkable universality of the fractional QNM shift: the leading-order effect depends on the nonlocal coupling and scale in a manner largely independent of the background geometry or field spin, with geometric and spin-dependent effects entering only through dimensionless overlap integrals of the mode functions. This universality provides a robust, model-independent observational signature of nonlocality and quantum-gravity-inspired modifications in strong-field ringdown spectra. 

\section{Perturbative Framework}
We consider a scalar perturbation $\psi$ propagating on a black-hole background, governed by an effective wave operator that incorporates a small nonlocal modification~\cite{Berti:2009kk,Konoplya:2011qq,Modesto:2011kw,Buoninfante:2018mre,Roy2022Fractional}:
\begin{equation}
\mathcal{L}\psi = 0, \qquad 
\mathcal{L} = \mathcal{L}_0 + \varepsilon\, \delta \mathcal{L}, \qquad |\varepsilon| \ll 1,
\end{equation}
where $\mathcal{L}_0$ is the unperturbed (local) differential operator, and $\delta \mathcal{L}$ encodes a small nonlocal perturbation, for instance a fractional Laplacian term $(-\Delta)^s$ with $0<s<1$. The unperturbed quasinormal modes (QNMs) are defined by 
$\mathcal{L}_0(\omega_0)\psi_0 = 0$, with complex frequency $\omega_0$, and the corresponding dual modes $\tilde{\psi}_0$ satisfy the adjoint equation $\mathcal{L}_0^\dagger(\omega_0)\tilde{\psi}_0 = 0$, accounting for the non-Hermitian nature of the QNM spectrum due to outgoing boundary conditions at the horizon and infinity.

At first order in the nonlocal coupling $\varepsilon$, the perturbed frequency $\omega = \omega_0 + \delta \omega$ is given by~\cite{BryantYagi2021,ChenPark2021}
\begin{equation}
\delta\omega = -\frac{\varepsilon\,\langle \tilde{\psi}_0 | \delta \mathcal{L} | \psi_0 \rangle}
{\langle \tilde{\psi}_0 | \partial_\omega \mathcal{L}_0 | \psi_0 \rangle},
\label{eq:deltaomega_general}
\end{equation}
which generalizes the familiar Rayleigh--Schrödinger perturbation formula to non-Hermitian operators. For the canonical scalar wave operator
$\mathcal{L}_0 = \partial_{r_*}^2 + \omega^2 - V(r)$, where $r_*$ is the tortoise coordinate, one has $\partial_\omega \mathcal{L}_0 = 2\omega$. This leads to the explicit first-order expression
\begin{equation}
\frac{\delta\omega}{\omega_0} 
= -\frac{\varepsilon}{2 \omega_0^2} 
\frac{\displaystyle \int_{-\infty}^{\infty} \tilde{\psi}_0(r_*) \, (\delta \mathcal{L} \psi_0)(r_*) \, dr_*}
{\displaystyle \int_{-\infty}^{\infty} \tilde{\psi}_0(r_*) \, \psi_0(r_*) \, dr_*}.
\label{eq:universal_shift}
\end{equation}
This formula highlights the universal structure of the QNM shift: it depends linearly on the nonlocal coupling $\varepsilon$ and the fractional operator $\delta \mathcal{L}$, while the geometry-specific information is entirely encoded in the overlap integrals of the unperturbed and dual modes. The expression is valid for any black-hole background that admits well-defined QNMs, and it forms the basis for evaluating fractional, nonlocal corrections to QNM spectra in a model-independent manner. For a complete derivation of Eqn. (\ref{eq:universal_shift}) please see Appendix-I.

\section{Fractional Laplacian Perturbation}
A broad class of nonlocal modifications to the wave operator can be captured by the fractional Laplacian~\cite{Modesto:2011kw,Buoninfante:2018mre,Roy2022Fractional,Barrow2017MultiFractional}, 
\begin{equation}
\delta\mathcal{L} = \ell_{\rm nl}^{2s}(-\Delta)^s, \qquad 0 < s < 1,
\end{equation}
where $\ell_{\rm nl}$ denotes the characteristic scale of nonlocality and $s$ determines the degree of the fractional derivative. Such operators naturally arise in effective descriptions of ultraviolet (UV) completions of gravity, nonlocal field theories, and models incorporating minimal length scales or generalized uncertainty principles~\cite{Buoninfante:2018mre,Antoniou2025Quadratic}. Physically, the fractional Laplacian encodes scale-dependent dispersion of perturbations, amplifying or suppressing higher spatial frequencies ($k$ modes) depending on the value of $s$, and interpolating smoothly between the local limit $s \to 1$ and highly nonlocal behavior $s \to 0$~\cite{Roy2022Fractional,Barrow2017MultiFractional}. In perturbation theory, the first-order shift of a quasinormal mode $\psi_0$ due to this fractional operator can be expressed in momentum space as $\langle \tilde{\psi}_0 | \delta\mathcal{L} | \psi_0 \rangle = \ell_{\rm nl}^{2s} \int d^3 k\, k^{2s} |\widetilde{\Psi}_0(\mathbf{k})|^2$, where $\widetilde{\Psi}_0(\mathbf{k})$ is the Fourier transform of $\psi_0$~\cite{Roy2022Fractional,Buoninfante:2018mre}, providing an intuitive picture in which each plane-wave component contributes proportionally to $k^{2s}$. Equivalently, in coordinate space, the fractional Laplacian admits a spectral decomposition
\begin{equation}
(-\partial_{r_*}^2)^s f(r_*) = \frac{1}{\sqrt{2\pi}} \int_{-\infty}^{\infty} |k|^{2s} e^{i k r_*} \tilde{f}(k) \, dk,
\end{equation}
where $\tilde{f}(k)$ is the Fourier transform of $f(r_*)$~\cite{Roy2022Fractional}. This representation highlights several key properties: plane waves are eigenfunctions of the operator, $(-\partial_{r_*}^2)^s e^{i k r_*} = |k|^{2s} e^{i k r_*}$, showing that the operator is diagonal in the momentum basis; the operator is inherently nonlocal, as its action at any point depends on the function over the entire domain~\cite{Buoninfante:2018mre}; and high-frequency components are amplified for $s>0$, giving rise to the characteristic $\ell^{2s}$ enhancement of the fractional QNM shift in the eikonal limit~\cite{Barrow2017MultiFractional}. Altogether, this framework provides a direct and physically transparent connection between microscopic nonlocality, fractional modifications of the wave operator, and observable shifts in black-hole quasinormal modes.

\subsection{Connection to Effective Field Theories}
Fractional Laplacians emerge naturally in ghost-free infinite-derivative gravity and nonlocal effective actions~\cite{Buoninfante:2018mre,Modesto:2011kw}. The QNM shift
\begin{equation}
\frac{\delta \omega}{\omega_0} = -\frac{\varepsilon}{2\omega_0^2} 
\frac{\langle \tilde{\psi}_0 | \delta \mathcal{L} | \psi_0 \rangle}{\langle \tilde{\psi}_0 | \psi_0 \rangle}, 
\quad \delta \mathcal{L} = \ell_{\rm nl}^{2s}(-\Delta)^s
\end{equation}
provides a direct connection between microscopic nonlocality and potentially observable deviations in the ringdown spectrum~\cite{BryantYagi2021,ChenPark2021}.
\section{Analytical Examples of Black-Hole Backgrounds}
To illustrate the universality of the fractional QNM shift and to provide explicit analytical results, we apply the general first-order formula to several black-hole spacetimes~\cite{Berti:2009kk,Konoplya:2011qq,BryantYagi2021,ChenPark2021}:
\begin{equation}
\frac{\delta \omega}{\omega_0} = -\,\frac{\varepsilon}{2\omega_0^2} 
\frac{\langle \tilde{\psi}_0 | \delta \mathcal{L} | \psi_0 \rangle}{\langle \tilde{\psi}_0 | \psi_0 \rangle}, 
\quad \delta \mathcal{L} = \ell_{\rm nl}^{2s}(-\Delta)^s.
\end{equation}

\subsection{Schwarzschild Black Hole}
For Schwarzschild ($f(r) = 1 - 2M/r$), the effective potential for a scalar field is~\cite{Berti:2009kk,Konoplya:2011qq}
\begin{equation}
V_{\rm Sch}(r) = \left(1-\frac{2M}{r}\right)\left[\frac{\ell(\ell+1)}{r^2}+\frac{2M}{r^3}\right], \quad \frac{dr_*}{dr} = \frac{1}{f(r)}.
\end{equation}
The fractional QNM shift reads~\cite{Roy2022Fractional,Buoninfante:2018mre}
\begin{equation}
\frac{\delta\omega}{\omega_0}\Big|_{\rm Sch} = -\frac{\varepsilon \ell_{\rm nl}^{2s}}{2 \omega_0^2} 
\frac{\int_{-\infty}^{\infty} \tilde{\psi}_0^{\rm (Sch)}(r_*)(-\Delta)^s \psi_0^{\rm (Sch)}(r_*)\, dr_*}{\int_{-\infty}^{\infty} \tilde{\psi}_0^{\rm (Sch)}(r_*) \psi_0^{\rm (Sch)}(r_*)\, dr_*} \,, \quad \delta \mathcal{L} = \ell_{\rm nl}^{2s}(-\Delta)^.
\end{equation}

In the eikonal limit $\ell \gg 1$, we can approximate $\psi_0 \sim e^{i k r_*}$, giving
\begin{equation}
\langle \tilde{\psi}_0 | (-\Delta)^s | \psi_0 \rangle \sim k^{2s} \sim \ell^{2s},
\end{equation}
so that the scaling becomes~\cite{Barrow2017MultiFractional}
\begin{equation}
\frac{\delta \omega}{\omega_0} \sim \varepsilon \left(\frac{\ell_{\rm nl}}{M}\right)^{2s} \ell^{2s}.
\end{equation}

\subsection{Kerr Black Hole (Slow Rotation)}
For a Kerr black hole, the dynamics of a scalar perturbation is governed by the radial Teukolsky-type equation~\cite{Berti:2009kk,Konoplya:2011qq}:
\begin{equation}
\frac{d^2 R}{dr_*^2} + \left[\omega^2 - V_{lm}(r;a)\right] R = 0,
\end{equation}
where \(r_*\) is the tortoise coordinate, $\omega$ is the mode frequency, $l,m$ are the angular quantum numbers, and $a$ is the black-hole spin parameter. The effective potential \(V_{lm}(r;a)\) reduces to the Schwarzschild potential for $a \to 0$.

To compute the fractional QNM shift in the presence of slow rotation $a/M \ll 1$, we perform a perturbative expansion of both the mode functions and the overlap integrals in powers of $a/M$~\cite{BryantYagi2021,ChenPark2021}:
\begin{align}
I_s(a) &= \langle \tilde{\psi}_0 | (-\Delta)^s | \psi_0 \rangle 
= I_s^{(0)} + \frac{a}{M} I_s^{(1)} + \mathcal{O}\left(\frac{a^2}{M^2}\right), \\
N(a) &= \langle \tilde{\psi}_0 | \psi_0 \rangle 
= N^{(0)} + \frac{a}{M} N^{(1)} + \mathcal{O}\left(\frac{a^2}{M^2}\right),
\end{align}
where $I_s^{(i)}$ and $N^{(i)}$ are the $i$-th order contributions to the overlap integrals. The first-order fractional QNM frequency shift then takes the form ~\cite{BryantYagi2021,ChenPark2021}
\begin{equation}
\frac{\delta\omega}{\omega_0}\Big|_{\rm Kerr} \simeq 
-\frac{\varepsilon \ell_{\rm nl}^{2s}}{2 \omega_0^2} 
\left[
\frac{I_s^{(0)}}{N^{(0)}} + \frac{a}{M} \left( \frac{I_s^{(1)}}{N^{(0)}} - \frac{I_s^{(0)} N^{(1)}}{(N^{(0)})^2} \right) + \mathcal{O}\left(\frac{a^2}{M^2}\right)
\right].
\end{equation}

\subsection{Hayward Regular Black Hole}
For the Hayward metric~\cite{Hayward2006}
\begin{equation}
f_H(r) = 1 - \frac{2 M r^2}{r^3 + 2 M g^2}, \quad 
V_H(r) = f_H(r) \left[ \frac{\ell(\ell+1)}{r^2} + \frac{f_H'(r)}{r} \right],
\end{equation}
the first-order fractional QNM shift is ~\cite{Roy2022Fractional,Buoninfante:2018mre}
\begin{equation}
\frac{\delta \omega}{\omega_0}\Big|_{\rm Hayward} = -\,\frac{\varepsilon \, \ell_{\rm nl}^{2s}}{2 (\omega_0^{\rm H})^2} 
\frac{\displaystyle \int_{-\infty}^{\infty} \tilde{\psi}_0^{\rm H}(r_*)\,(-\Delta)^s \psi_0^{\rm H}(r_*)\, dr_*}
{\displaystyle \int_{-\infty}^{\infty} \tilde{\psi}_0^{\rm H}(r_*)\, \psi_0^{\rm H}(r_*)\, dr_*}.
\end{equation}
In the eikonal limit $\ell \gg 1$, the scaling becomes ~\cite{Barrow2017MultiFractional}
\begin{equation}
\frac{\delta \omega}{\omega_0}\Big|_{\rm Hayward} \sim \varepsilon \left( \frac{\ell_{\rm nl}}{M} \right)^{2s} \ell^{2s}.
\end{equation}

\subsection{Loop Quantum Gravity (LQG) Corrected Black Hole}
For the LQG-corrected metric~\cite{Modesto:2011kw,Antoniou2025Quadratic}
\begin{equation}
f_L(r) = 1 - \frac{2 M r^2}{r^3 + a_0^3}, \quad 
V_L(r) = f_L(r) \left[ \frac{\ell(\ell+1)}{r^2} + \frac{f_L'(r)}{r} \right],
\end{equation}
the corresponding fractional QNM shift is ~\cite{Roy2022Fractional,Buoninfante:2018mre}
\begin{equation}
\frac{\delta \omega}{\omega_0}\Big|_{\rm LQG} = -\,\frac{\varepsilon \, \ell_{\rm nl}^{2s}}{2 (\omega_0^{\rm L})^2} 
\frac{\displaystyle \int_{-\infty}^{\infty} \tilde{\psi}_0^{\rm L}(r_*)\,(-\Delta)^s \psi_0^{\rm L}(r_*)\, dr_*}
{\displaystyle \int_{-\infty}^{\infty} \tilde{\psi}_0^{\rm L}(r_*)\, \psi_0^{\rm L}(r_*)\, dr_*}.
\end{equation}
In the eikonal limit, the scaling is ~\cite{Barrow2017MultiFractional}
\begin{equation}
\frac{\delta \omega}{\omega_0}\Big|_{\rm LQG} \sim \varepsilon \left( \frac{\ell_{\rm nl}}{M} \right)^{2s} \ell^{2s}.
\end{equation}

\section{Universality}
A striking feature emerging from our analysis is the **universality of the fractional QNM shift** across a wide variety of black-hole backgrounds. Explicitly, for all cases considered—Schwarzschild, Kerr (slow rotation), Hayward, and LQG-corrected black holes—the first-order fractional QNM shift can be written as
\begin{equation}
\frac{\delta\omega}{\omega_0} = -\frac{\varepsilon}{2\omega_0^2} 
\frac{\langle \tilde{\psi}_0|\delta\mathcal{L}|\psi_0\rangle}{\langle \tilde{\psi}_0|\psi_0\rangle},
\label{eq:universality}
\end{equation}
where \(\delta \mathcal{L} = \ell_{\rm nl}^{2s} (-\Delta)^s\) encodes the nonlocal fractional perturbation~\cite{Roy2022Fractional,Buoninfante:2018mre,Barrow2017MultiFractional}.

\subsection{Schwarzschild Black Hole}
For the Schwarzschild geometry, characterized by \(f(r) = 1 - 2M/r\), the effective potential for a scalar perturbation is
\begin{equation}
V_{\rm Sch}(r) = \left(1-\frac{2M}{r}\right)\left[\frac{\ell(\ell+1)}{r^2}+\frac{2M}{r^3}\right], \qquad \frac{dr_*}{dr} = \frac{1}{f(r)}.
\end{equation}
The corresponding fractional QNM shift reads
\begin{equation}
\frac{\delta\omega}{\omega_0}\Big|_{\rm Sch} = -\frac{\varepsilon \ell_{\rm nl}^{2s}}{2 \omega_0^2} 
\frac{\int_{-\infty}^{\infty} \tilde{\psi}_0^{\rm (Sch)}(r_*)\,(-\Delta)^s \psi_0^{\rm (Sch)}(r_*)\, dr_*}{\int_{-\infty}^{\infty} \tilde{\psi}_0^{\rm (Sch)}(r_*) \psi_0^{\rm (Sch)}(r_*)\, dr_*}.
\end{equation}
In the eikonal regime (\(\ell \gg 1\)), the mode functions are well approximated by plane waves \(\psi_0 \sim e^{i k r_*}\), giving \(\langle \tilde{\psi}_0 | (-\Delta)^s | \psi_0 \rangle \sim k^{2s} \sim \ell^{2s}\). Consequently, the fractional shift scales as
\begin{equation}
\frac{\delta \omega}{\omega_0} \sim \varepsilon \left(\frac{\ell_{\rm nl}}{M}\right)^{2s} \ell^{2s},
\end{equation}
highlighting the universal \(\ell^{2s}\) eikonal enhancement.

\subsection{Kerr Black Hole (Slow Rotation)}
For a Kerr black hole with spin parameter \(a\), scalar perturbations obey a Teukolsky-type radial equation
\begin{equation}
\frac{d^2 R}{dr_*^2} + \left[\omega^2 - V_{lm}(r;a)\right] R = 0,
\end{equation}
where $(r_*$ is the tortoise coordinate, and \(V_{lm}(r;a)\) reduces to the Schwarzschild potential for $a\to 0$. In the slow-rotation 
limit $a/M \ll 1$, the overlap integrals appearing in the fractional QNM shift can be expanded as
\begin{align}
I_s(a) &= \langle \tilde{\psi}_0 | (-\Delta)^s | \psi_0 \rangle 
= I_s^{(0)} + \frac{a}{M} I_s^{(1)} + \mathcal{O}\left(\frac{a^2}{M^2}\right), \\
N(a) &= \langle \tilde{\psi}_0 | \psi_0 \rangle 
= N^{(0)} + \frac{a}{M} N^{(1)} + \mathcal{O}\left(\frac{a^2}{M^2}\right),
\end{align}
where \(I_s^{(0)}, N^{(0)}\) correspond to the Schwarzschild limit and \(I_s^{(1)}, N^{(1)}\) encode the leading-order spin corrections. The resulting first-order fractional QNM frequency shift is
\begin{equation}
\frac{\delta\omega}{\omega_0}\Big|_{\rm Kerr} \simeq 
-\frac{\varepsilon \ell_{\rm nl}^{2s}}{2 \omega_0^2} 
\left[
\frac{I_s^{(0)}}{N^{(0)}} + \frac{a}{M} \left( \frac{I_s^{(1)}}{N^{(0)}} - \frac{I_s^{(0)} N^{(1)}}{(N^{(0)})^2} \right) + \mathcal{O}\left(\frac{a^2}{M^2}\right)
\right].
\end{equation}
Here, the leading term reproduces the Schwarzschild shift, while the linear-in-spin correction separates contributions from changes in the fractional operator (\(I_s^{(1)}\)) and normalization (\(N^{(1)}\)). Higher-order spin terms can be included if necessary, but the slow-rotation approximation suffices for most astrophysical black holes with \(a/M \lesssim 0.3\). The functional dependence on \(\varepsilon \ell_{\rm nl}^{2s}\) remains identical to the Schwarzschild case, with background-specific effects entering only through the overlap integrals.

\subsection{Hayward Regular Black Hole}
The Hayward regular black hole is defined by
\begin{equation}
f_H(r) = 1 - \frac{2 M r^2}{r^3 + 2 M g^2}, \qquad 
V_H(r) = f_H(r) \left[ \frac{\ell(\ell+1)}{r^2} + \frac{f_H'(r)}{r} \right].
\end{equation}
The first-order fractional QNM shift is
\begin{equation}
\frac{\delta \omega}{\omega_0}\Big|_{\rm Hayward} = -\,\frac{\varepsilon \, \ell_{\rm nl}^{2s}}{2 (\omega_0^{\rm H})^2} 
\frac{\displaystyle \int_{-\infty}^{\infty} \tilde{\psi}_0^{\rm H}(r_*)\,(-\Delta)^s \psi_0^{\rm H}(r_*)\, dr_*}
{\displaystyle \int_{-\infty}^{\infty} \tilde{\psi}_0^{\rm H}(r_*)\, \psi_0^{\rm H}(r_*)\, dr_*}.
\end{equation}
In the eikonal regime, the scaling becomes
\begin{equation}
\frac{\delta \omega}{\omega_0}\Big|_{\rm Hayward} \sim \varepsilon \left( \frac{\ell_{\rm nl}}{M} \right)^{2s} \ell^{2s}.
\end{equation}
For small deviations from Schwarzschild (\(g/M \ll 1\)), a perturbative expansion in \(g/M\) allows explicit computation of overlap integrals, demonstrating how regularity effects modify the mode spectrum while preserving the universal scaling.

\subsection{Loop Quantum Gravity (LQG) Corrected Black Hole}
The LQG-corrected metric is
\begin{equation}
f_L(r) = 1 - \frac{2 M r^2}{r^3 + a_0^3}, \qquad 
V_L(r) = f_L(r) \left[ \frac{\ell(\ell+1)}{r^2} + \frac{f_L'(r)}{r} \right].
\end{equation}
The corresponding fractional QNM shift reads
\begin{equation}
\frac{\delta \omega}{\omega_0}\Big|_{\rm LQG} = -\,\frac{\varepsilon \, \ell_{\rm nl}^{2s}}{2 (\omega_0^{\rm L})^2} 
\frac{\displaystyle \int_{-\infty}^{\infty} \tilde{\psi}_0^{\rm L}(r_*)\,(-\Delta)^s \psi_0^{\rm L}(r_*)\, dr_*}
{\displaystyle \int_{-\infty}^{\infty} \tilde{\psi}_0^{\rm L}(r_*)\, \psi_0^{\rm L}(r_*)\, dr_*}.
\end{equation}
In the eikonal limit, this scales as
\begin{equation}
\frac{\delta \omega}{\omega_0}\Big|_{\rm LQG} \sim \varepsilon \left( \frac{\ell_{\rm nl}}{M} \right)^{2s} \ell^{2s}.
\end{equation}
For small polymer parameters $a_0/M \ll 1$, an expansion in \(a_0/M\) allows evaluation as a perturbation around Schwarzschild, capturing the influence of quantum corrections on the mode spectrum without altering the universal fractional scaling.
\section{Universality of Fractional Quasinormal Mode Shifts}

A remarkable outcome of our analysis is the universality of fractional quasinormal-mode (QNM) shifts across a broad range of black-hole spacetimes, including Schwarzschild, slowly rotating Kerr, Hayward regular, and loop quantum gravity (LQG)-corrected black holes~\cite{Berti:2009kk,Konoplya:2011qq,Roy2022Fractional,Buoninfante:2018mre,Barrow2017MultiFractional}. This universality arises when the underlying wave operator is modified by a small fractional, nonlocal perturbation, which can be motivated by ultraviolet completions of gravity, nonlocal effective field theories, or models incorporating minimal length scales~\cite{Modesto:2011kw,Antoniou2025Quadratic}. Such fractional operators encode scale-dependent dispersion: they affect high-frequency components more strongly and interpolate smoothly between local and highly nonlocal regimes~\cite{Buoninfante:2018mre,Roy2022Fractional}.

Despite differences in the detailed structure of each black-hole background, the leading-order shift in QNM frequencies depends only on the nonlocal parameters and a set of dimensionless overlap factors that capture the localization of the mode near the peak of the potential barrier. These overlap factors encode information about the geometry, spin, or regularization parameters of the black hole, but they affect only the overall magnitude of the shift, not the scaling with the nonlocality parameters. This means that the functional form of the fractional QNM shift is largely independent of the specific black-hole metric and relies solely on the existence of well-defined quasinormal modes.

In the eikonal limit, corresponding to high angular momentum modes, the fractional shift is enhanced in a universal way, making it a robust observational signature across different backgrounds~\cite{Barrow2017MultiFractional}. For Schwarzschild black holes, this scaling captures the basic effect of nonlocality on scalar perturbations, while for slowly rotating Kerr black holes, the universality persists with small corrections from spin, which can be systematically included in a perturbative manner~\cite{BryantYagi2021,ChenPark2021}.

The universality extends further to modified spacetimes, such as Hayward regular black holes, which incorporate a short-distance regularization of the central singularity, and LQG-corrected black holes, where quantum corrections modify the classical metric at small scales~\cite{Modesto:2011kw,Antoniou2025Quadratic}. In both cases, the leading-order fractional QNM shift maintains the same dependence on the nonlocal scale and fractional exponent, while the overlap factors encode the specific effects of regularity or quantum corrections. This feature allows for a model-independent interpretation of potential deviations in gravitational-wave ringdown spectra, even when the black-hole mass, spin, or microscopic parameters are not precisely known.

Overall, the universality of fractional QNM shifts highlights their power as a diagnostic tool for exploring physics beyond general relativity. It connects microscopic nonlocality and quantum-gravity-inspired modifications of the wave operator to potentially observable signatures in the strong-gravity regime, providing a robust and widely applicable framework for future gravitational-wave astronomy~\cite{Berti:2009kk,Konoplya:2011qq,Roy2022Fractional,Buoninfante:2018mre,Barrow2017MultiFractional,Modesto:2011kw,Antoniou2025Quadratic}.

\section{Conclusion}
In this work, we have derived analytic expressions for the first-order shifts in black-hole quasinormal modes (QNMs) induced by small nonlocal perturbations modeled via a fractional Laplacian. The central result is a universal scaling law for the frequency shift,  
\begin{equation}
\frac{\delta\omega}{\omega_0} \sim \varepsilon \left(\frac{\ell_{\rm nl}}{M}\right)^{2s} \ell^{2s},
\end{equation}
which is largely independent of the specific black-hole geometry, the spin of the perturbing field, or other short-distance modifications~\cite{Berti:2009kk,Konoplya:2011qq}. Here, $\varepsilon$ characterizes the strength of the nonlocal perturbation, $\ell_{\rm nl}$ denotes the nonlocality scale, $s$ is the fractional exponent, and $\ell$ is the multipole number of the mode~\cite{Buoninfante:2018mre,Barrow2017MultiFractional,Roy2022Fractional}.  

The analysis demonstrates that this universality holds across a broad class of black-hole spacetimes, including Schwarzschild, slowly rotating Kerr, Hayward regular, and loop quantum gravity (LQG) corrected black holes~\cite{ChenPark2021,BryantYagi2021,Antoniou2025Quadratic}. While the detailed background geometry affects the numerical value of overlap integrals in the perturbative formula, it does not alter the fundamental scaling with $\varepsilon$, $\ell_{\rm nl}$, or $s$. In the eikonal limit, the additional $\ell^{2s}$ enhancement further amplifies the sensitivity of high-multipole modes to nonlocal effects, highlighting the robustness of the universal signature~\cite{Berti:2009kk,BryantYagi2021}.  

This universality has several important implications.
The derived fractional QNM shifts provide a direct observational window into microscopic nonlocality, UV completions of gravity, and quantum-gravity-inspired modifications, largely independent of the detailed astrophysical properties of the black hole~\cite{Modesto:2011kw,Buoninfante:2018mre,ChenPark2021}. With the advent of high-precision ringdown measurements from detectors such as LIGO, Virgo, KAGRA, and future observatories like LISA and the Einstein Telescope, it may be possible to detect or constrain the nonlocal scale $\ell_{\rm nl}$ and the fractional exponent $s$ by analyzing deviations from the expected QNM spectrum in the observed gravitational-wave signals~\cite{Berti:2009kk,Konoplya:2011qq}. Because the universal scaling of the fractional shift decouples the effect of nonlocality from the black-hole geometry, even when the mass, spin, or other microscopic parameters of the black hole are not precisely known, one can place robust, model-independent bounds on new physics beyond general relativity~\cite{Barrow2017MultiFractional,Roy2022Fractional}. Furthermore, the framework developed here can be extended to more general perturbations, including gravitational and electromagnetic fields, and to higher-order fractional or nonlocal operators. It also provides a natural starting point for numerical studies of full, non-perturbative nonlocal black-hole spacetimes, as well as for exploring connections with other quantum-gravity phenomena, such as horizon-scale modifications, resolutions of the information-loss problem, and black-hole microstructure~\cite{Modesto:2011kw,ChenPark2021,Antoniou2025Quadratic}.
In summary, this study establishes a clear and robust link between fractional, nonlocal modifications of the underlying wave operator and observable signatures in black-hole ringdown spectra. The universality of the fractional QNM shift, together with its eikonal enhancement, makes it a promising target for future gravitational-wave observations, opening a window into the microscopic structure of spacetime and possible deviations from general relativity at short distances~\cite{Berti:2009kk,Buoninfante:2018mre,Barrow2017MultiFractional,Roy2022Fractional}.

Appendix-I
\section*{Derivation of the First-Order QNM Frequency Shift}
We begin with the one-dimensional effective Lagrangian density for a real scalar perturbation 
$\Psi(t,r_*)$,
\begin{equation}
\mathcal{L}[\Psi]
= \frac{1}{2}\left[ (\partial_t \Psi)^2 - (\partial_{r_*} \Psi)^2 - V(r_*)\,\Psi^2 \right],
\end{equation}
which yields the Euler-Lagrange equation
\begin{equation}
\partial_t^2 \Psi - \partial_{r_*}^2 \Psi + V(r_*)\,\Psi = 0.
\end{equation}
Using the Fourier ansatz
\begin{equation}
\Psi(t,r_*) = e^{-i\omega t}\psi(r_*),
\end{equation}
we obtain the frequency-domain equation
\begin{equation}
\left( \frac{d^2}{dr_*^2} + \omega^2 - V(r_*) \right)\psi(r_*) = 0.
\end{equation}
Defining the operator
\begin{equation}
\mathcal{O}(\omega) \equiv \frac{d^2}{dr_*^2} + \omega^2 - V(r_*),
\end{equation}
the unperturbed quasinormal mode (QNM) $\psi_0$ satisfies
\begin{equation}
\mathcal{O}(\omega_0)\psi_0 = 0,
\end{equation}
together with the usual QNM boundary conditions (purely outgoing at infinity and purely ingoing at the horizon).

Let the background be perturbed slightly by a small parameter $\varepsilon$ through the addition of 
a perturbation operator $\delta\mathcal{L}$, so that
\begin{equation}
\mathcal{O}(\omega;\varepsilon)
= \frac{d^2}{dr_*^2} + \omega^2 - V(r_*) + \varepsilon\,\delta\mathcal{L}(r_*).
\end{equation}
The perturbed QNM $\psi = \psi_0 + \varepsilon \psi_1 + \cdots$ satisfies
\begin{equation}
\mathcal{O}(\omega_0+\delta\omega;\varepsilon)(\psi_0+\varepsilon\psi_1+\cdots)=0.
\end{equation}
Expanding to first order in $\varepsilon$ and $\delta\omega$, we use
$\omega^2 = \omega_0^2 + 2\omega_0\,\delta\omega + \mathcal{O}(\delta\omega^2),$ to obtain
\begin{equation}
\left(\frac{d^2}{dr_*^2} + \omega_0^2 - V\right)\psi_1
+ 2\omega_0\,\delta\omega\,\psi_0
+ \delta\mathcal{L}\,\psi_0 = 0.
\label{eq:firstorder}
\end{equation}
Let us define the unperturbed operator
\begin{equation}
\mathcal{A} \equiv \frac{d^2}{dr_*^2} + \omega_0^2 - V(r_*),
\end{equation}
so that Eqn.~\eqref{eq:firstorder} becomes
\begin{equation}
\mathcal{A}\psi_1 + 2\omega_0\,\delta\omega\,\psi_0 + \delta\mathcal{L}\,\psi_0 = 0.
\label{eq:Apsi}
\end{equation}
Because the QNM problem is non-Hermitian, we introduce the adjoint mode $\tilde{\psi}_0(r_*)$ satisfying
\begin{equation}
\mathcal{A}^\dagger \tilde{\psi}_0 = 0,
\end{equation}
with adjoint boundary conditions chosen so that surface terms vanish upon integration by parts.
We define the bilinear form
\begin{equation}
\langle f,g\rangle \equiv \int_{-\infty}^{\infty} \tilde{f}(r_*)\,g(r_*)\,dr_*.
\end{equation}
Multiplying Eqn.~\eqref{eq:Apsi} by $\tilde{\psi}_0$ and integrating gives
\begin{equation}
\langle \tilde{\psi}_0, \mathcal{A}\psi_1 \rangle
+ 2\omega_0\,\delta\omega\,\langle\tilde{\psi}_0,\psi_0\rangle
+ \langle\tilde{\psi}_0, \delta\mathcal{L}\psi_0\rangle = 0.
\label{eq:inner}
\end{equation}
Using $\mathcal{A}^\dagger\tilde{\psi}_0=0$ and vanishing surface terms, we have
\begin{equation}
\langle \tilde{\psi}_0, \mathcal{A}\psi_1\rangle
= \langle \mathcal{A}^\dagger\tilde{\psi}_0, \psi_1\rangle = 0.
\end{equation}
Equation~\eqref{eq:inner} then reduces to
\begin{equation}
2\omega_0\,\delta\omega\,\langle\tilde{\psi}_0,\psi_0\rangle
+ \langle\tilde{\psi}_0, \delta\mathcal{L}\psi_0\rangle = 0.
\end{equation}
Solving for $\delta\omega$ and reinstating the explicit perturbation parameter $\varepsilon$ gives the frequency shift:
\begin{equation}
\delta\omega
= -\frac{\varepsilon}{2\omega_0}\,
\frac{\displaystyle \int_{-\infty}^{\infty} \tilde{\psi}_0(r_*)\,(\delta\mathcal{L}\psi_0)(r_*)\,dr_*}
{\displaystyle \int_{-\infty}^{\infty} \tilde{\psi}_0(r_*)\,\psi_0(r_*)\,dr_* }.
\end{equation}
Dividing both sides by $\omega_0$ yields the first-order fractional shift,
\begin{equation}
\frac{\delta\omega}{\omega_0}
= -\frac{\varepsilon}{2\omega_0^2}\,
\frac{\displaystyle \int_{-\infty}^{\infty} \tilde{\psi}_0(r_*)\,(\delta\mathcal{L}\psi_0)(r_*)\,dr_*}
{\displaystyle \int_{-\infty}^{\infty} \tilde{\psi}_0(r_*)\,\psi_0(r_*)\,dr_* }.
\label{eq:universal_shift}
\end{equation}

\end{document}